# FINDING SIGNS OF LIFE ON TRANSITING EARTH-LIKE PLANETS: HIGH-RESOLUTION TRANSMISSION SPECTRA OF EARTH THROUGH TIME AROUND FGKM HOST STARS

Lisa Kaltenegger[1, 2], Zifan Lin[1,2] & Sarah Rugheimer[1,3]
[1]Carl Sagan Institute, Space Science Building 311, Ithaca, NY 14850, USA
[2]Cornell University, Astronomy and Space Sciences Building, Ithaca, NY 14850, USA
[3]Oxford University, Oxford, UK

ABSTRACT
   The search for life in the universe mainly uses modern Earth as a template. However, we know that Earth's atmospheric composition changed significantly through its geological evolution. Recent discoveries show that transiting, potentially Earth-like exoplanets orbit a wide range of host stars, which strongly influence their atmospheric composition and remotely detectable spectra. Thus, a database for transiting terrestrial exoplanet around different host stars at different geological times, is a crucial missing ingredient to support observational searches for signs of life in exoplanet atmospheres.

   Here, we present the first high-resolution transmission spectra database for Earth-like planets, orbiting a wide range of host stars, throughout four representative stages of Earth's history. These correspond to a prebiotic high $CO_2$-world - about 3.9 billion years ago in Earth's history - and three epochs through the rise of oxygen from 0.2% to modern atmospheric levels of 21%. We demonstrate that the spectral biosignature pairs $O_2$ + $CH_4$ and $O_3$ + $CH_4$ in the atmosphere of a transiting Earth-like planet would show a remote observer that a biosphere exists for oxygen concentrations of about 1% modern Earth's - corresponding to about 1 to 2 billion years ago in Earth's history - for all host stars.

   The full model and high-resolution transmission spectra database, covering 0.4μm to 20μm, for transiting exoplanets - from young prebiotic worlds to modern Earths-analogs - orbiting a wide range of host stars is available online. It can be used as a tool to plan and optimize our observation strategy, train retrieval methods, and interpret upcoming observations with ground- and space-based telescopes.

**Introduction**

The discovery of thousands of exoplanets has uncovered the wide range of both stellar hosts, as well as ages of potential Earth-like planets (e.g. Kaltenegger & Sasselov 2011, Kane et al. 2016, Berger et al. 2019; Johns et al. 2018). The diversity keeps increasing with new discoveries by NASA's TESS mission (see e.g. Quintana et al. 2014, Kaltenegger et al. 2019a) and ground-based searches like MEarth, SPECULOOS, and CARMENES (e.g. Nutzman, & Charbonneau 2008, Guillon et al. 2017, Luque et al. 2019, Kaltenegger et al. 2019b). Future ground-based Extremely Large Telescopes (ELTs) as well as JWST are designed to obtain the first measurements of the atmospheric composition of Earth-sized planets (see e.g. Snellen et al. 2013; Rodler & Lopez-Morales 2014; Ben-Amin et al. 2018, Serindag & Snellen 2019, Batalha et al. 2019).

While we expect a large variety of rocky planets with alternative evolutionary paths from Earth, our planet's atmosphere, which has evolved significantly since formation (see, e.g., Zahnle et al. 2007; Lyons et al. 2014), provides the only template of a habitable rocky planet's evolution to date.

Previous work by the authors investigated the changes in the reflection and emission spectra of Earth-like planets through geological time orbiting different host stars based on atmospheric models of Earth through its geological history from anoxic to modern Earth (for details see Kaltenegger et al. 2007, Rugheimer & Kaltenegger 2018). These models showed that a remote observer could identify a biosphere for about 2 billion years in Earth's history in emission and reflection. The spectra presented in those papers provide templates for observation of directly imaged worlds as well as planets seen around secondary eclipse, as "Pale Blue Dots". The spectra of a transiting Earth-analog planet through its geological evolution orbiting a Sun-analog star has been modeled and discussed in Kaltenegger et al. (2020).

However, small transiting exoplanets have been found around a wide range of host stars. Because of the change in the emitted stellar energy distribution (SED) of a star with surface temperature, the host's irradiation has a strong influence on the photochemistry of a rocky planet's atmosphere and its resulting chemical composition (see e.g. Kasting et al. 1993, Segura et al. 2005, Rauer et al. 2011, Hedelt et al. 2013, Rugheimer et al. 2013, Kaltenegger 2017). To explore the effect of the host star's irradiation on Earth-like planetary atmospheres the authors previously modeled such evolving planets from prebiotic to modern Earth atmospheres around an extensive grid of F0V to M8V stars ($T_{eff}$ 7000–2400K) (see details in Rugheimer et al. 2013, 2015), using Exo-Prime a 1D climate-photochemistry-Radiative Transfer model (see e.g. Rugheimer & Kaltenegger 2018 for details).

Here, we use these atmosphere models to generate the high-resolution transmission database of spectra of Earth-like planets orbiting a wide range of Sun-like host stars at different stages in their evolution covering the visible to infrared wavelength range (0.4 to 20μm). Our transit spectra include climate indicators like $H_2O$ and $CO_2$ as well as biosignatures, remotely detectable atmospheric gases that are produced by life and are not readily mimicked by abiotic processes, like the $CH_4+O_2$ and $CH_4+O_3$ pairs (Lederberg 1965, Lovelock 1965). On short timescales, the two species react to produce $CO_2$ and $H_2O$ and therefore, must be constantly replenished to maintain detectable concentrations (see e.g. review Kaltenegger 2017). Sagan et al. (1993) analyzed the Galileo probe emergent spectrum of Earth and concluded that the large amount of $O_2$ in the presence of $CH_4$ is strongly suggestive of biology, as Lovelock (1965) and Lederberg (1965) had suggested earlier. It is their quantities and detection along with other atmospheric species in the planetary context that solidify a biological origin (as discussed in detail in several recent reviews e.g. Kasting et al. 2014, Kaltenegger 2017, Fuji et al. 2018, Schwieterman et al. 2018).

In this letter, we present a database of transmission spectra from the visible to the infrared for evolving Earth-like planets orbiting a wide range of Sun-like host stars and identify at what point a remote observer can identify biosignatures in their atmospheres. Section 2 describes the atmospheric composition of terrestrial planets throughout their evolution for a wide range of host stars and discusses the high-resolution transit spectra simulations. Section 3 presents our results and section 4 discusses our results.

Our high-resolution transit spectra database for evolving planets orbiting a wide range of host stars is freely available online at (carlsaganinstitute.org/data, and DOI: 10.5281/zenodo.4029370) as a tool to prioritize targets, prepare effective observation strategies and to guide first interpretation of atmospheric spectra of Earth-like planets, using our planet's evolution as a template.

2. Methods

The four geological epochs we model correspond to a young, prebiotic, high $CO_2$-world about 3.9 billion years ago (Ga) in Earth's history, and 3 epochs through the rise of oxygen from 0.2% oxygen, about 1 to 2 billion years ago in Earth's history, to modern atmospheric levels of 21%.

The $CO_2$-rich atmosphere (epoch 1) represents a prebiotic world around 3.9Ga. A Paleo- and Meso-proterozoic Earth (about 2 to 1Ga), when oxygen started to rise in Earth's atmosphere is represented in epoch 2 using 0.21% $O_2$ (1% present atmospheric level (PAL)). Neoproterozoic Earth (about 0.8 to 0.5Ga) (epoch 3) represents a time period when the oxygen concentration had risen to about 10% PAL (2.1% $O_2$) for Earth, corresponds to the proliferation of multicellular life. Modern Earth is modeled with 21% $O_2$ (epoch 4). The time ranges given in Table 1 for epoch 3 and epoch 4 (see also Kaltenegger et al. 2020) represent the shift to later times for the rise of oxygen in Earth's evolution (see e.g. Lyones et al. 2014), instead of geological times given in our earlier paper (Kaltenegger et al. 2007), which based $O_2$ concentrations on work by Holland et al. (2006).

## 2. 1 Atmospheric models

The models for each epoch are discussed in detail in Rugheimer & Kaltenegger (2018) and summarized in Table 1. Temperature and atmospheric mixing ratios from the model paper are summarized in Fig. 1.

All our models assume Earth-radius and -mass. All epochs assume a 1 bar surface pressure, consistent with geological evidence for paleo-pressures close to modern values or lower (e.g. Som et al. 2012; Marty et al. 2013).

Our model planet receives similar irradiation as Earth did during its evolution at 1AU around an evolving Sun. For the Sun we use an evolution model (Claire et al. 2012) while for other host stars, we reduce the overall luminosity by a factor corresponding to the evolution of the Sun (for details see Rugheimer & Kaltenegger 2018). This approach does not account for increased UV activity for young M stars. Note that this procedure is not meant to represent consistent stellar evolution across FGKM stars (as explained in detail in Rugheimer & Kaltenegger 2018). Our models compare planets with the same incident bolometric flux for a wide range of Sun-like host stars for planets with atmospheric compositions and biogenic flux modeled using our Earth's evolution.

The reduction in incident flux for models of earlier epochs around our host star grid also corresponds to planetary models with increased orbital distances from the host stars, a topic we plan to explore in future work. Across host stellar types, the surface UV environment is similar to Earth's through geological time (see Rugheimer & Kaltenegger 2018, O'Malley-James & Kaltenegger 2017) and not a major concern for surface habitability (O'Malley-James & Kaltenegger 2019).

## 2.2 Transmission Spectra models

We model all high-resolution transmission spectra of planets at a resolution of $0.01 cm^{-1}$ for Earth-like planets orbiting different host stars for 4 geological epochs with EXO-Prime (for details see Rugheimer & Kaltenegger 2018). The transmission spectra in Fig. 2 and Fig. 4 are smoothed with a triangular kernel to a resolving power of 300 for clarity.

All high-resolution spectra can be downloaded from our online database (Resolution $\lambda/\Delta\lambda > 100,000$ from 0.4 and 10µm, $\lambda/\Delta\lambda > 50,000$ from 10µm to 20µm).

EXO-Prime is a 1D iterative climate-photochemistry atmospheric model coupled with a line-by-line Radiative Transfer code developed for Earth (e.g. Traub & Stier 1976; Traub & Jucks 2002, Chance & Traub 2007) and adapted to rocky exoplanets (see e.g. Kaltenegger et al. 2007). EXO-Prime has been validated from the visible to infrared through comparison to Earth seen as an exoplanet by missions like the Mars Global Surveyor, EPOXI, multiple earthshine observations and Shuttle data (Kaltenegger et al. 2007, Kaltenegger & Traub 2009; Rugheimer et al. 2013). We use opacities from the 2016 HITRAN database (Gordon et al. 2017) for $O_2$, $O_3$, OH, O, $H_2O$, $HO_2$, $H_2O_2$, $CO_2$, CO, $H_2CO$, $CH_4$, $CH_3O_2$, $CH_3OOH$, $CH_3Cl$, HCl, HOCl, $Cl_2O_2$, ClO, $ClONO_2$, $SO_2$, $H_2S$, $H_2SO_4$, HSO, HS, $H_2$, H, $N_2O$, $NO_2$, $NO_3$, NO, $HNO_2$ and $HNO_3$. $N_2O_5$ (Wagner & Birk 2003) (Sharpe et al. 2004) is included through cross-sections. We use measured continua data instead of line-by-line calculations in the far wings for $N_2$, $H_2O$ and $CO_2$ (see Traub & Jucks 2002) and include $CO_2$ line mixing (see also Niro et al. 2005).

Refraction in deeper atmospheric regions of a planet's atmosphere can deflect light away from a distant observer for Earth-like planets (see e.g. Sidis et al. 2010, Betremieux & Kaltenegger 2014, Robinson et al. 2017), which sets the lowest level Earth can be probed to in primary transit to about 12km (e.g. Betremieux & Kaltenegger 2014). For other host stars the depth varies from F0V 15.7km to 0km for M8V host stars (F0V 15.7km, F7V 13.8km, G2V 12.7km, G8V 11.7km, K2V 9.6km, K7V 6.6km, M1A 3.9km, M3A 1.7km, M8A 0km) (following Betremieux & Kaltenegger 2014). Clouds do not significantly affect the strengths of the spectral features in Earth's transmission because most clouds on Earth are located at altitudes below 12km. However, for planets, which can be probed deeper than 12km, clouds will start to obscure part of the spectra. Clouds in transmission spectra will obscure the spectral features below the cloud layer, if they occur close or on the terminator region that is probed during primary transit (e.g., Seager et al. 2005, Kaltenegger et al. 2009, Robinson et al. 2011; Betremieux & Kaltenegger 2014). Currently there is no conclusion on the location of clouds on Earth-like planets orbiting different host stars. Therefore, we have indicated the modern Earth's cloud heights at 1km, 6km and 12km height (following Kaltenegger et al. 2007) in Fig.2 and Fig.3 as dashed lines.

We assume full primary transit geometry for our spectra. The dominant contributions to transmission spectra from Earth-like planet atmospheres come from the atmosphere below 60km in the wavelength range modeled ($10^{-4}$ bar). However, to fully capture all spectral lines in the high-resolution spectra, we extend the atmosphere modeled in Rugheimer & Kaltenegger (2018) isothermal above that region to $10^{-7}$ bar. This choice is driven by the largest features, $CO_2$ at 4.3μm. Note that outside the considered wavelength range, higher parts of the atmosphere can contribute significantly to the transmission spectra, e.g. up to 180km for modern Earth's UV transmission spectrum (see Betremieux & Kaltenegger 2013). No noise has been added to the transmission spectra to provide theoretical input spectra for any instrument.

## 3. Results

Transmission spectra from the visible to infrared (0.4μm to 20μm) for Earth-like planets orbiting a wide range of Sun-like stars show a range of spectral features. We shortly describe the change in atmospheric species due to the influence of the host star for the different geological epochs modeled, which maps onto the transmission spectra. We focus on climate indicators $H_2O$ and $CO_2$ as well as the biosignature pairs $CH_4 + O_2$ and $CH_4 + O_3$ (Lederberg 1965, Lovelock 1965), which is our most robust sign of life in the atmosphere on other planets. Other gases, which could indicate biology but also have alternative explanations, include Nitrous oxide, $N_2O$, and chloromethane, $CH_3Cl$ (see e.g. Segura et al. 2005; Kasting et al. 2014, Kaltenegger 2017, Schwieterman 2018, Grenfell 2017). Note that our models assume biological $N_2O$ sources for all biotic atmospheres (see Rugheimer & Kaltenegger 2018).

Fig.2 shows the corresponding spectra and star to planet contrast ratio for our models for four host stars, representing our full grid, at a resolution of 300 with the most prominent spectral features identified. The contrast ratio varies by about two orders of magnitude between our grid stars due to their difference in size, making planets orbiting small stars promising targets for biosignature detection. Several features overlap at the resolution of $\lambda/\Delta\lambda=300$, which is shown in Fig.2 for clarity, and are not specifically labeled. In the high-resolution online transmission spectra individual spectral lines can be easily discerned for these molecules, as shown in Fig.3 for $O_2$ and $O_3$ for biotic atmospheres (epoch 2 to 4), for a resolution of $\lambda/\Delta\lambda > 100,000$ as proposed for several instruments on the ELTs. Note that for planets around cooler stars than our Sun, which can be probed below 12km, clouds will influence the detectability of any spectral features. However, with no clear answer where such clouds will form, we indicate the three Earth cloud levels as dashed lines in Fig.2 and Fig.3 and show the transmission spectra down to the refraction limit.

All high-resolution spectra are available online at carlsaganinstitute.org/data, and DOI: 10.5281/zenodo.4029370.

### 3.1 Transmission spectra of Earth-like planets through geological time

The transmission spectra show large changes of the atmospheric composition from a prebiotic world to modern Earth (Fig.2).

Throughout the atmospheric evolution of our models, different absorption features dominate Earth's spectrum with $CO_2$ being dominant in Early Earth models, where it is more abundant and $O_2$ and $O_3$ features increasing with $O_2$ abundance from a Paleo- and Meso-proterozoic Earth to modern Earth (Table 1).

For modern Earth Fig.2 shows strong absorption features of $O_2$ at 0.76μm, with a weaker feature at 0.69μm, and $O_3$ at 9.6μm, with a weaker feature at 9μm, as well as a broad $O_3$ triangular feature from 0.45μm to 0.74μm (the Chappuis band). These features become weaker with decreasing oxygen content.

$CH_4$ features increasing with strength for younger biotic Earth models (7.7μm, 3.3μm, 2.32μm, 1.69μm, 1.00μm, 0.89μm, 0.73, and 0.6μm). Note that the prebiotic Earth model assumes low $CH_4$ concentrations.

$H_2O$ abundance and absorption feature strength increase with increasing surface temperature and consequence evaporation rate and shows spectral features at a wide range of wavelength, the strongest are labeled in Fig.2 at 17-20μm, 6.4μm, 2.6μm, 1.88μm, 1.41μm, 1.13μm, 0.94μm, 0.82μm and 0.72μm. With increasing $CO_2$ abundance for earlier Earth models, $CO_2$ spectral feature at 15μm, 10.4μm, 9.3μm, 2.7μm, 2.35μm, 2.03μm, 1.62μm, 1.59μm, 1.21μm and 1.06μm become stronger.

$N_2O$ shows absorption features at 3.7, 4.5, 7.8, 8.6, 10.65 and 17μm. $CH_3Cl$ shows features at 3.35μm, 7μm, 9.7μm, and 13.7μm. Feature strength decreases with decreasing concentrations for younger Earth models. $N_2O$ and $CH_3Cl$ features are not labeled in Fig.2 because they overlap with $CO_2$, $H_2O$ and $CH_4$ features and can thus not be easily identified in low resolution spectra: e.g. the 17μm $N_2O$ and the 13.7μm $CH_3Cl$ feature are located in the wings of the 15μm $CO_2$ feature.

### 3.2 Transmission spectra of Earth-like planets across host star type

Fig.2 shows that within each atmospheric model (see Rugheimer & Kaltenegger 2018 for details), increasing UV flux photolyzes larger amounts of $CH_4$ and $H_2O$ in the stratosphere, cooling it.

In modern atmospheres, heating due to $O_3$ dominates and thus higher stellar UV flux leads to hotter stratospheric temperatures. $O_3$ is primarily produced by wavelengths below 2,400Å and destroyed by wavelengths below 3,200Å, therefore the UV incident flux shortward and longward of 2,400Å sets the abundance of atmospheric $O_3$ (see e.g. Segura et al. 2005; Domagal-Goldman et al. 2011), increasing the ozone column densities for cooler host stars in our sample. Note that due to photolysis of other species such as $CO_2$ and $H_2O$ in the prebiotic atmosphere at 3.9Ga, there is enough free oxygen to form $O_3$ in the stratosphere, although surface concentrations are up to 10 orders of magnitude lower (see Rugheimer & Kaltenegger 2018 for details).

The mixing ratio of $CH_4$ is given in Table 1, with the highest $CH_4$ mixing ratio at epoch 2, assuming increased $CH_4$ productions by methanogens. $N_2O$ concentrations in our models assume increasing biological flux for later epochs and no $N_2O$ flux for the prebiotic Earth model. For higher UV radiation from the host star, $N_2O$ is destroyed by photolysis. $H_2O$ increases with planetary surface temperature and subsequent evaporation.

For all epochs and host stars, $H_2O$ and $CH_4$ show absorption features, which overlap in the visible wavelength range (Fig.3). While the overlap of the features will make it challenging to distinguish $H_2O$ and $CH_4$ in the visible at low resolution, the increase of the methane features in the visible for younger Earth models increases our opportunity to discover the biosignature pair $O_2$ and $CH_4$ in the visible wavelength range. Methane and water features can be distinguished in the visible by measuring the flux at 1.7μm, where only methane absorbs, but water does not.

### 3.3. High-resolution spectra increase detectability of Biosignatures

As shown in Fig.2 the observability of biosignature generally increases with planetary age because the detectability of oxygen and ozone as part of the $CH_4$ oxygen/ozone biosignature pair increases. Fig.4 shows the change in both the $O_2$ feature at 0.76μm and the $O_3$ feature at 9.6μm through geological time for four different host stars for a minimum resolution of 100,000 between 0.4 and 10μm and a minimum of 50,000 between 10 to 20μm. We do not show the $CH_4$ features in high resolution here, because it can be seen for all biotic epochs at 7.7μm and 2.3μm in low resolution (Fig.2).

The models show the rise of oxygen from 0.01 PAL for a Paleo- and Meso-proterozoic Earth to 21% oxygen for modern Earth. In addition to the change in the abundance of oxygen and ozone as described, Fig.3 shows how the refraction cutoff in the lower atmosphere influences the spectra. It varies from 15.7km (F0) to 0km (M8). The contrast ratio varies by about two orders of magnitude between our grid stars due to their difference in size, making planets orbiting small stars promising targets for biosignature detection.

The abundance of $CH_4$ increases for younger biotic atmosphere models. Therefore, the biosignature pair $CH_4$ + $O_2$ could also become observable in the visible for younger Earth models, especially in high resolution, where water and methane features can be distinguished. As shown in Kaltenegger et al. (2020) for Earth, a biosphere could be detected for 0.1 PAL oxygen, corresponding to about 1 - 2 billion years ago in Earth's history (e.g. Lyons et al. 2014). The observability of the spectral $O_2$, $O_3$ and $CH_4$ features for lower concentrations depends on the resolution as well as the signal to noise ratio of the specific instrument chosen.

### 4. Summary & Discussion

The diversity of known transiting exoplanets suggests that we will encounter a large diversity of terrestrial exoplanet atmospheres. To expand from the modern Earth models commonly used, our high-resolution transmission spectra database presented here provides a range of contrast ratios of spectral features from the visible to the infrared for transiting Earth-like exoplanets orbiting a wide range of Sun-like host stars ($T_{eff}$ 7,000–2,400K) – from young prebiotic worlds to modern Earths-analogs.

We chose atmospheres representative of four geological epochs of Earth's history: A prebiotic high $CO_2$-world about 3.9 billion years ago in Earth's history (epoch 1), as well as 3 epochs through the rise of $O_2$ from 0.2% to present atmospheric levels, which started about 2.4 billion years ago on Earth.

Different absorption features dominate the terrestrial planets' transmission spectra with $CO_2$ features dominating in Early Earth models, where they are more abundant. $O_2$ and $O_3$ spectral features become stronger with increasing abundance during the rise of oxygen (epoch 2 to 4). All spectra are available in high resolution online (carlsaganinstitute.org/data, and DOI: 10.5281/zenodo.4029370).

### 4.1. When could signs of life be detected in the atmosphere of Earth-like transiting planets?

Using our own planet's evolution as our basis, we created a high-resolution transmission spectra database of Earth-like planets of different incident flux orbiting a wide range of Sun-like stars to explore when signs of life could be detected in their atmospheres.

The transmission spectra show spectral features, which would indicate a biosphere on exoplanets orbiting F0 to M8 stars for 0.1PAL of modern Earth's oxygen, corresponding to about 1 to 2 billion years ago in Earth's history.

In addition, we show that the increasing strength of methane features in the visible transmission spectra for younger Earth models would significantly improve the detectability of the biosignature pair $O_2+CH_4$ in the visible wavelength range. For modern Earth the methane features in the visible range are very small and challenging to assess. However, note that most visible $CH_4$ features overlap with water features and have to be distinguished (Fig.3). This can be done at specific wavelengths where $CH_4$ features do not overlap with other chemicals like at 1.7 or 2.3μm or in high resolution.

### 4.2. Atmosphere evolution for Earth and Earth-like planets

We do not yet understand what sets the evolution timescale on a terrestrial planet. Because no self-consistent evolution models exist for all host stars, we reduce the overall luminosity of all stellar types by the same amount, based on an evolutionary model for the Sun (Claire et al. 2012) and the resulting incident flux Earth received at each geological epoch modeled. This does not represent similar timescales for the grid of host stars, with the evolution of the hotter F stars being much faster than the cooler grid stars.

The reduction in stellar incident flux for models of earlier epochs also corresponds to planetary models with increased orbital distances from the host stars, a topic we plan to explore in future work. Meanwhile the models of early epochs around the different host stars show specific examples of the influence of orbital distance on planetary transit spectra. However, our models compare planets that receive the same bolometric incident flux across stellar hosts with specific planetary models corresponding to Earth's evolution.

Therefore, while this spectral database provides limited insights into the effect of distance from the host star in the habitable Zone due to the reduced stellar flux at early epochs corresponding to a larger orbital distance of a planet as well, it compares similar incident flux, not similar orbital distances. In addition, note that the refraction limits which sets how deep a planet's atmosphere can be probed, would change with orbital distances from the host star (see e.g. Betremieux & Kaltenegger.2014, Robinson et al. 2017). In our model spectra the refraction limit is set for the 1AU equivalent distance for all planet models.

### 4.3 Biosignatures need to be set in context of their star and planet

We chose the $CH_4 +O_2$ (Lederberg.1965, Lovelock.1965) and $CH_4+O_3$ pairs as the most reliable biosignatures for Earth-like planets (see e.g. reviews: Kaltenegger 2017, Grenfell 2017). On short timescales, the two species react to produce $CO_2$ and $H_2O$ and therefore, must be constantly replenished to maintain detectable concentrations. Neither gas alone is a reliable biosignature.

$CH_4$ can be produced abiotically e.g. in volcanic eruptions or serpentinization. Several mechanisms can produce atmospheric oxygen and ozone in a terrestrial planet's atmosphere through photolysis of $H_2O$ or $CO_2$, for planets e.g. with low-$H_2$/high-$CO_2$ atmospheres (Domagal-Goldman et al. 2014), with low-pressure atmospheres (Wordsworth & Pierrehumbert 2014) or $CO_2$-rich atmospheres orbiting M dwarfs (Hu et al. 2012; Tian et al. 2014). These results reinforce the caution to not use an individual spectral feature like $O_2$ alone as a biosignature.

As photolysis depends strongly on the UV radiation, understanding the host star and the radiation environment of the planet is important to interpret biosignature detections (see e.g. Airapetian et al. 2020, France et al. 2016, Rimmer & Rugheimer 2019).

Absorption features for the biosignature pairs $CH_4+O_2$ and $CH_4+O_3$ for epoch 2 to 4 are shown in Fig.2 at several wavelengths. The strongest features in the modeled wavelength range for these gases in the visible to NIR are $O_3$ (0.4-0.6μm), $O_2$ (0.76μm), $CH_4$ (2.3μm), and in the thermal IR $O_3$ (9.6μm) and $CH_4$ (7.6μm).

The high resolution transmission spectra database for Earth-like exoplanets orbiting a wide range of Sun-like host stars - from young prebiotic worlds to modern Earths-analogs - is available online and can be used as a tool to optimize our observation strategy, train retrieval methods, as well as interpret upcoming observations with JWST as well as ground-based Extremely Large Telescopes and future mission concepts like Origins, HabEx, and LUVOIR.

With the next generation of telescopes on the ground and in space, we will, for the first time in human history, be able to characterize rocky planets in the Habitable Zone of their host stars. Drawing on the rich history of Earth - the only habitable world we know - increases our chances to discover signs of life in the universe.

ACKNOWLEDGEMENTS
The authors acknowledge funding from the Carl Sagan Institute and the Brinson Foundation. LK thanks Ryan MacDonald for insightful comments on the paper draft.

**Table 1:** Surface mixing ratios for major atmospheric gases in our model atmospheres for 4 epochs through Earth's geological history from prebiotic to anoxic atmospheres. The models are representative of a prebiotic world about 3.9 billion years ago in Earth's history (epoch 1), as well as 3 models which capture the rise of oxygen from a Neoproterozoic Earth modeled with 0.01 PAL $O_2$ to modern Earth with 21% $O_2$ (see Rugheimer & Kaltenegger 2018).

| Host star | Time (Ga) | Solar Const. | Ep | $T_{surf}$ | Surface Mixing ratio | | | | |
|---|---|---|---|---|---|---|---|---|---|
| | | | | | $CO_2$ | $CH_4$ | $O_2$ | $O_3$ | $N_2O$ |
| Sun | now | 1.00 | 4 | 288.1 | 3.55E-04 | 1.60E-06 | 2.10E-01 | 2.41E-08 | 3.00E-07 |
| F0 | now | 1.00 | 4 | 279.8 | 3.55E-04 | 3.40e-06 | 2.10E-01 | 5.00e-08 | 1.80e-07 |
| K7 | now | 1.00 | 4 | 300.2 | 3.55E-04 | 1.30e-04 | 2.10E-01 | 2.40e-08 | 8.60e-07 |
| M8 | now | 1.00 | 4 | 301.9 | 3.55E-04 | 1.00e-03 | 2.10E-01 | 1.60e-09 | 3.10e-06 |
| Sun | 0.5 - 0.8 | 0.95 | 3 | 302.2 | 1.00E-02 | 4.15E-04 | 2.10E-02 | 2.02E-08 | 9.15E-08 |
| F0 | 0.5 - 0.8 | 0.95 | 3 | 294.4 | 1.00E-02 | 5.60e-04 | 2.10E-02 | 3.40e-08 | 1.10e-07 |
| K7 | 0.5 - 0.8 | 0.95 | 3 | 305.7 | 1.00E-02 | 1.20e-02 | 2.10E-02 | 1.20e-08 | 1.60e-07 |
| M8 | 0.5 - 0.8 | 0.95 | 3 | 284.1 | 1.00E-02 | 1.10e-02 | 2.10E-02 | 2.90e-09 | 8.60e-07 |
| Sun | 1.0 - 2.0 | 0.87 | 2 | 291.1 | 1.00E-02 | 1.65E-03 | 2.10E-03 | 7.38E-09 | 8.37E-09 |
| F0 | 1.0 - 2.0 | 0.87 | 2 | 288.2 | 1.00E-02 | 1.20e-03 | 2.10E-03 | 2.40e-08 | 8.30e-09 |
| K7 | 1.0 - 2.0 | 0.87 | 2 | 288.0 | 1.00E-02 | 2.00e-02 | 2.10E-03 | 5.80e-09 | 4.90e-08 |
| M8 | 1.0 - 2.0 | 0.87 | 2 | 280.1 | 1.00E-02 | 4.00e-03 | 2.10E-03 | 2.00e-09 | 3.40e-07 |
| Sun | 3.9 | 0.75 | 1 | 290.2 | 1.00E-01 | 1.65E-06 | 1.00E-13 | 2.55E-19 | 0 |
| F0 | 3.9 | 0.75 | 1 | 277.3 | 1.00E-01 | 1.70e-06 | 1.00E-13 | 3.70e-19 | 0 |
| K7 | 3.9 | 0.75 | 1 | 297.4 | 1.00E-01 | 1.70e-06 | 1.00E-13 | 1.20e-19 | 0 |
| M8 | 3.9 | 0.75 | 1 | 300.3 | 1.00E-01 | 1.70e-06 | 1.00E-13 | 7.70e-19 | 0 |

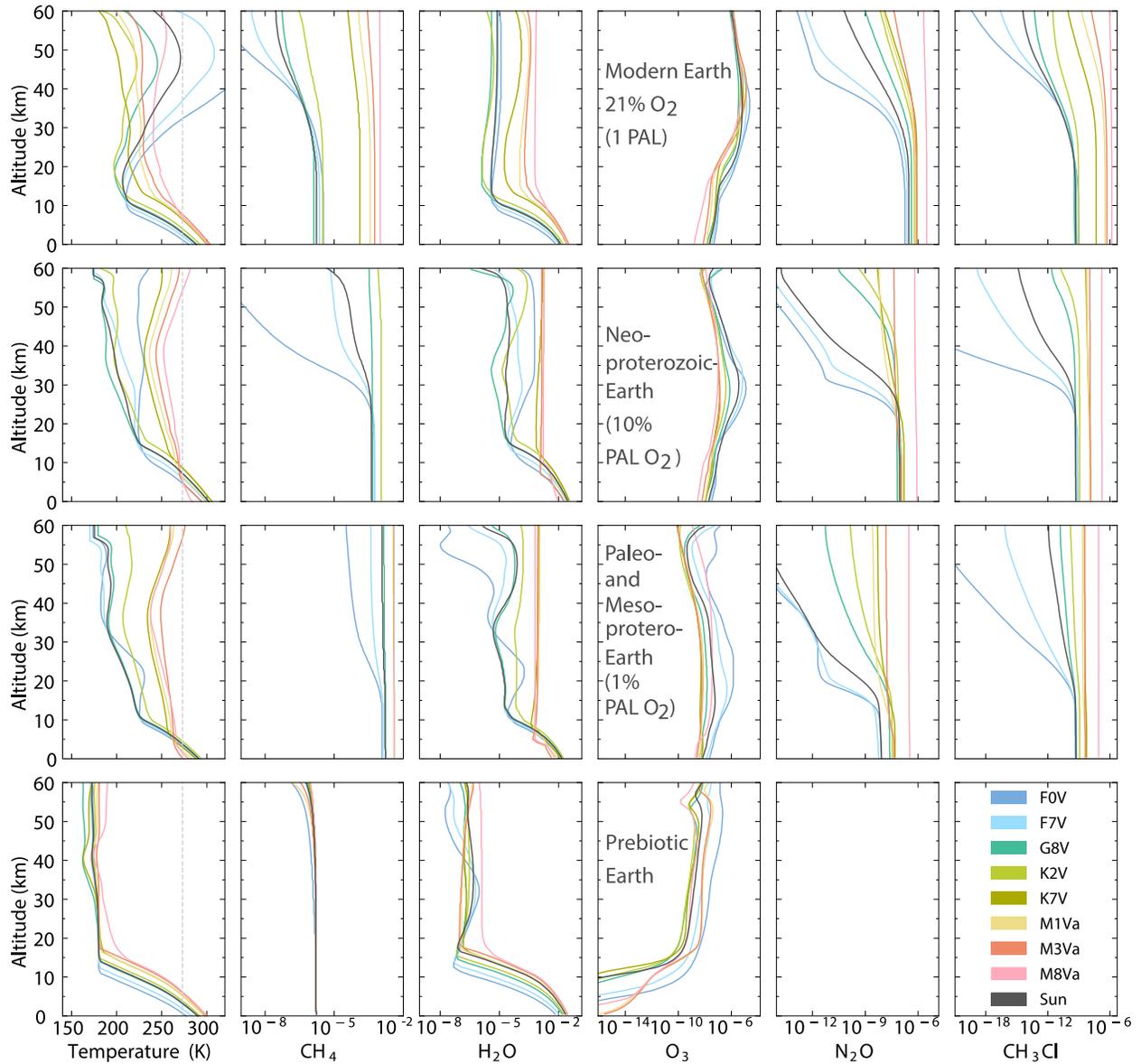

**Figure 1:** Temperature profile and mixing ratios for the major atmospheric gases in our atmospheric models for Earth-like planets around different host stars, representing 4 epochs through Earth's evolution from a prebiotic atmosphere around 3.9 billion years ago to modern Earth (see Rugheimer & Kaltenegger 2018 for details).

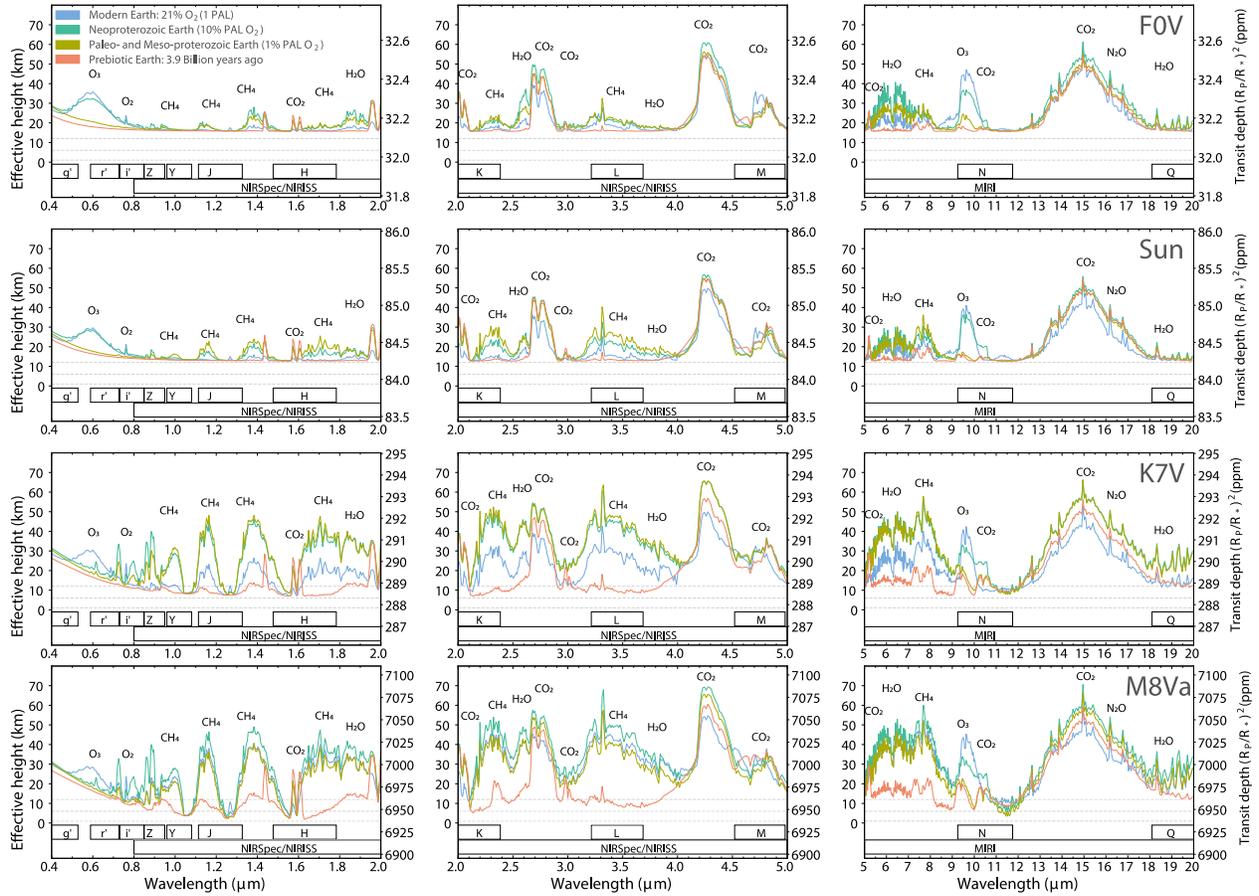

**Figure 2:** Model Spectra for Earth through geological time from 0.4 to 20 μm shown at a resolution of 300 for 4 epochs through Earth's geological time from an anoxic atmosphere 3.9 billions years ago to 3 models which capture the rise of oxygen from 0.01 PAL $O_2$ to 1PAL (21% $O_2$) on modern Earth, which started around 2.4 billion years ago. We show models for four host stars of our grid (F0, Sun, K7 and M8). Spectra for all grid stars are available online (carlsaganinstitute.org/data, and DOI: 10.5281/zenodo.4029370).

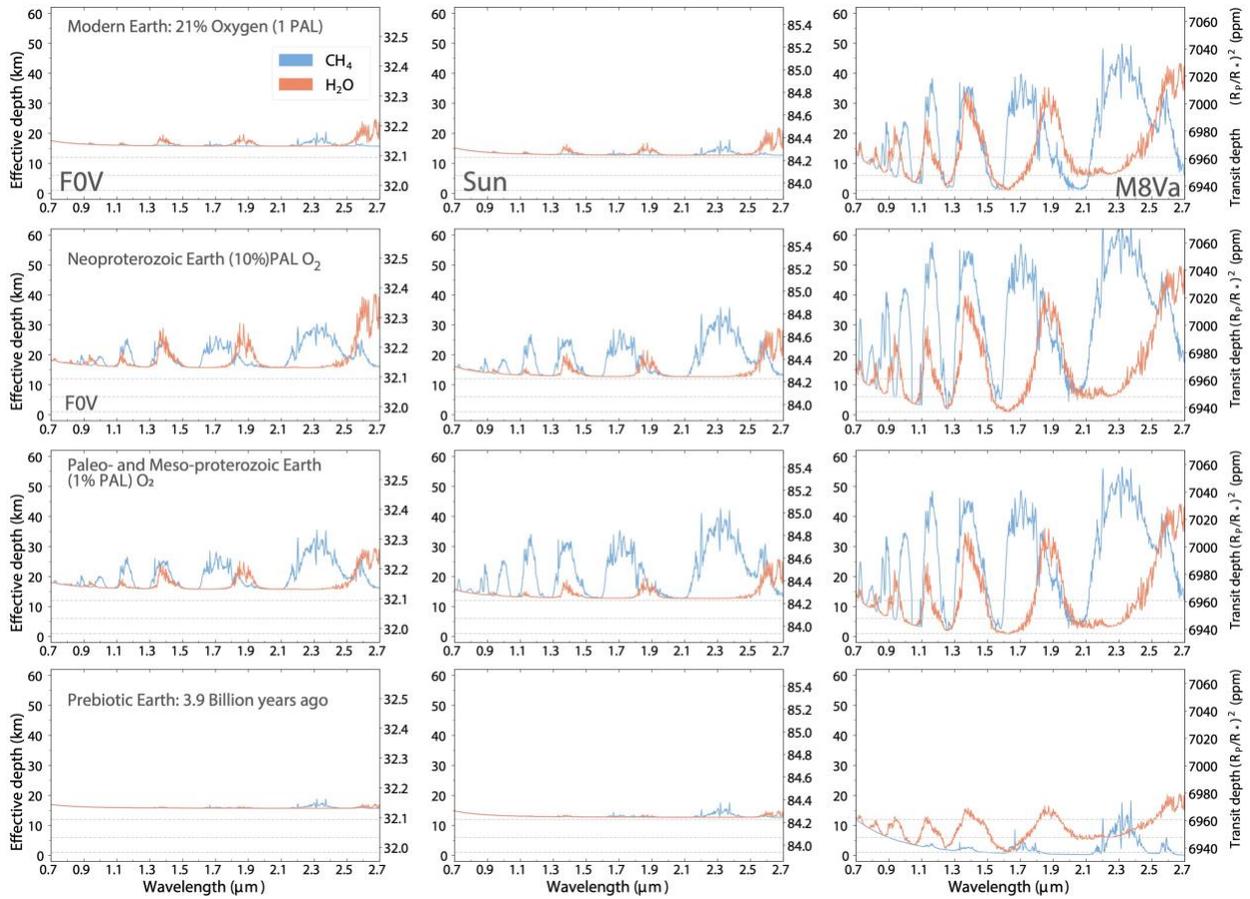

**Figure 3:** Overlapping water and methane absorption features in Earths visible transmission spectra through geological time from 0.7 to 2.7μm shown at a resolution of 300 for the 4 epochs modeled for three host stars in our grid (F0, Sun, M8). Spectra for all grid stars are available online (carlsaganinstitute.org/data, and DOI: 10.5281/zenodo.4029370).

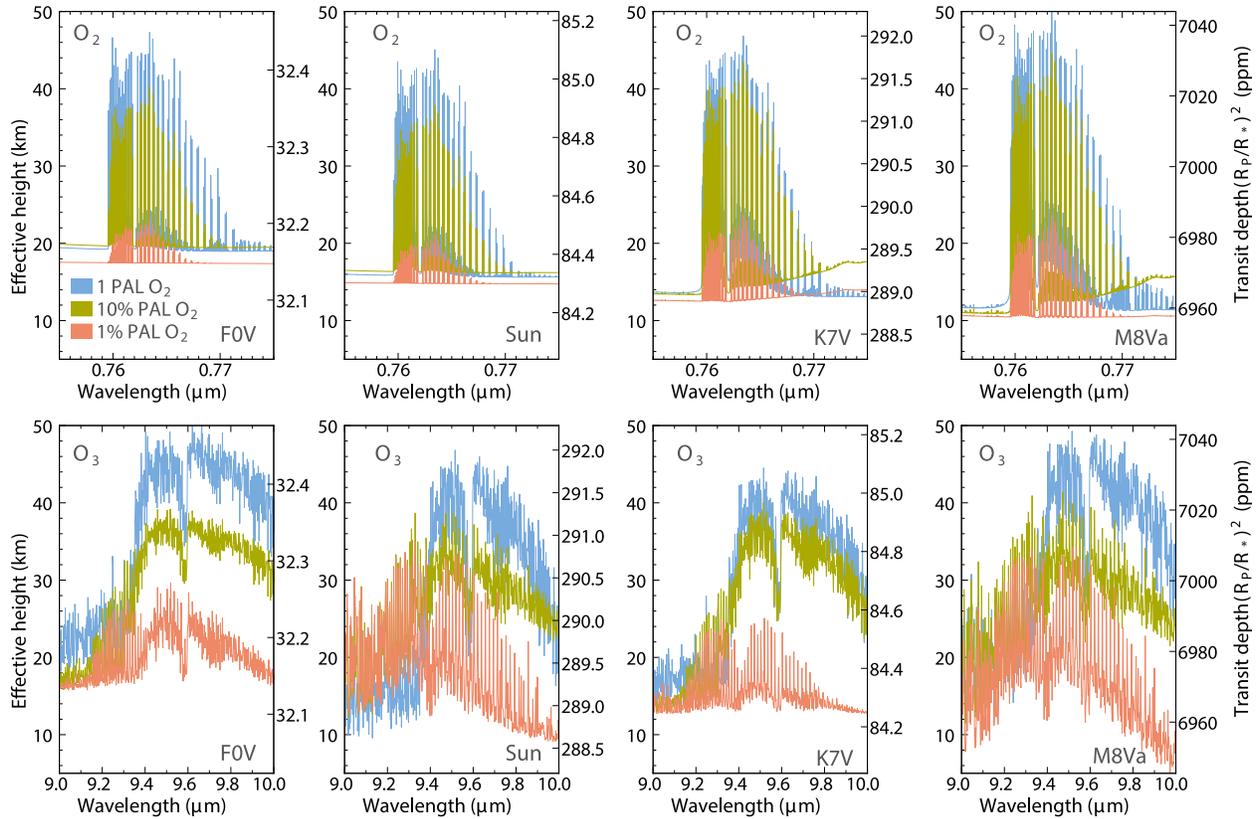

**Figure 4**: High-resolution (R > 100,000) for the 0.76μm $O_2$ and 9.6μm $O_3$ feature for the rise of oxygen from 0.01 to the present atmospheric level (PAL) of oxygen, which is 21% in Earth's modern atmosphere, for four host stars of our grid (F0, Sun, K7 and M8). Spectra for all grid stars are available online (carlsaganinstitute.org/data, and DOI: 10.5281/zenodo.4029370).

14